# Order picking efficiency: A scattered storage and clustered allocation strategy in automated drug dispensing systems

Mengge Yuan

Faculty of Science, Kunming University of Science and Technology

Kunming, Yunnan, China, 650500

yuanmengge0818@126.com

Ning Zhao

Faculty of Science, Kunming University of Science and Technology

Kunming, Yunnan, China, 650500

zhaoning@kust.edu.cn

Kan Wu*

Business Analytics Research Centre

Chang Gung University, Taoyuan City, Taiwan, 33302

kan626@gmail.com

Lulu Cheng

Faculty of Science, Kunming University of Science and Technology

Kunming, Yunnan, China, 650500

chenglulu0206@163.com

Correspondence information: Kan Wu, Business Analytics Research Centre,

Chang Gung University, Taoyuan City, Taiwan, 33302, kan626@gmail.com

# Order picking efficiency: A scattered storage and clustered allocation strategy in automated drug dispensing systems

A. Mengge Yuan[a], B. Ning Zhao[a], C. Kan Wu[b,1], D. Lulu Cheng[a]

[a] Faculty of Science, Kunming University of Science and Technology, Kunming, Yunnan, 650500, China

[b] Business Analytics Research Centre, Chang Gung University, Taoyuan City, 33302, Taiwan

**Abstract**

In the smart hospital, optimizing prescription order fulfilment processes in outpatient pharmacies is crucial. A promising device, automated drug dispensing systems (ADDSs), has emerged to streamline these processes. These systems involve human order pickers who are assisted by ADDSs. The ADDS's robotic arm transports bins from storage locations to the input/output (I/O) points, while the pharmacist sorts the requested drugs from the bins at the I/O points. This paper focuses on coordinating the ADDS and the pharmacists to optimize the order-picking strategy. Another critical aspect of order-picking systems is the storage location assignment problem (SLAP), which determines the allocation of drugs to storage locations. In this study, we consider the ADDS as a smart warehouse and propose a two-stage scattered storage and clustered allocation (SSCA) strategy to optimize the SLAP for ADDSs.

---

[1] Corresponding author. E-mail address: kan626@gmail.com

The first stage primarily adopts a scattered storage approach, and we develop a mathematical programming model to group drugs accordingly. In the second stage, we introduce a sequential alternating (SA) heuristic algorithm that takes into account the drug demand frequency and the correlation between drugs to cluster and locate them effectively. To evaluate the proposed SSCA strategy, we develop a double objective integer programming model for the order-picking problem in ADDSs to minimize the number of machines visited in prescription orders while maintaining the shortest average picking time of orders. The numerical results demonstrate that the proposed strategy can optimize the SLAP in ADDSs and improve significantly the order-picking efficiency of ADDSs in a human-robot cooperation environment.

# 1. Introduction

The emerging trend of smart pharmacies has gained significant attention in recent years. To improve the efficiency of smart pharmacy operations, intelligent facilities, such as automated drug dispensing systems (ADDSs), are introduced. Consequently, the focus of attention naturally gravitates towards the operational efficiency of ADDSs. There have been notable advancements in autonomous machines designed to execute essential warehousing functions (Azadeh et al., 2019). However, there are certain limitations to the fully automation of ADDS. The cooperation between humans and robots becomes crucial (Boysen et al., 2017), particularly concerning two critical aspects: the accurate sorting of prescription orders based on drug dosage and ensuring prescription precision. The basic elements of robot-assisted order picking are depicted in Fig. 1. Within this concept, humans (pharmacists) and the ADDS cooperate. In such a human-robot cooperation environment, it would be unwise to focus solely on robotic picking efficiency while disregarding the impact of humans.

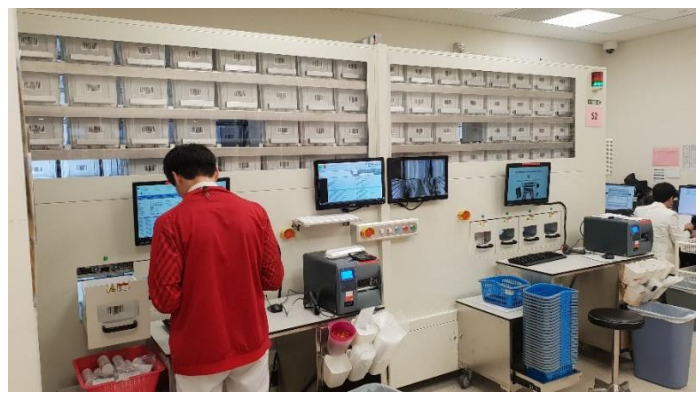

Fig. 1. The cooperation between humans and robots.

In the era of e-commerce, the need for responsive order fulfilment processes and

next- or even same-day deliveries has gained strategic significance across various industries (Boysen et al., 2019). To improve the efficiency of ADDSs, we focus on the prescription order-picking process which is the central activity of ADDSs. Order picking is generally recognized as the most expensive warehouse operation because it tends to be either very labour-intensive or very capital-intensive (Gu et al., 2007). Thus, even a slight improvement in the ordering picking process would bring a significant efficiency boost to the warehouse operation. Moreover, in the order-picking process travel time usually contributes more than half of its total time costs (John J. Bartholdi & Hackman, 2019). Similarly, in smart pharmacies, there is a high demand from patients for timely and accurate prescription order fulfilment.

The storage location assignment problem (SLAP), which determines how products are assigned to storage locations, plays a critical role in influencing order-picking efficiency. Hausman et al. (1976) and Roodbergen and Vis (2009) expounded on the general storage location assignment strategies, such as random storage, dedicated storage, class-based storage, and storage location assignment based on full turnover, and so on. However, these strategies overlook the correlation between products in historical orders, i.e., how frequently the two products are ordered together. Apart from the turnover speed needed, product correlation is another important attribute derived from order history. If this attribute is ignored in the assignment, two types of drugs that are frequently requested together (e.g. aspirin and acetaminophen) might be assigned to storage locations far from each other, which can unnecessarily increase prescription order picking time.

Besides the above traditional storage location assignment strategies, recent advancements have introduced innovative storage strategies aimed at minimizing unproductive travel distances and enhancing order-picking efficiency. To decrease the average distance from anywhere to the threshold, scattered storage is proposed (Weidinger & Boysen, 2018), in which the products are scattered all around the warehouse and the probability of always having some items per stock-keeping unit close by is increased. Scattered storage is one of the storage assignment strategies adopted by many warehouses and distribution centres of online retailers (Boysen et al., 2019). The scattered storage strategy focuses on the essential item-to-location correspondence and reduces the average travel time by balancing the distribution of the items from the same stock-keeping unit (Yang et al., 2020). The basic decision problems based on a scattered storage strategy are the same as in any other warehouse. As far as we know, there is currently no literature about the application of scattered storage strategy to ADDSs. This gap in research is the focus of our study.

By considering the correlation between the two products, it is possible to assign them to several groups. These groups then can be strategically allocated to the storage locations to minimize order picking time. The dominant approach in the literature for cluster-based allocation (Chuang et al., 2012 ; Mirzaei et al., 2021 ; Zhang, 2016) involves decomposing the problem of minimizing travel time into two key steps: grouping and allocation. In the first step, products are grouped based on their correlation, and in the second step, these grouped products are assigned to locations close to each other. Note that optimizing both problems in a decomposition approach does not guarantee an overall optimal solution.

The ADDS operates as a small-scale automated storage and retrieval system and has both drug storage and automatic dispensing functions. It can be viewed as a new type of warehouse that is designed for prescription order fulfilment. In this system, each drug can be stored in more than one location (bin), but each location (bin) can store only one type of drug. A drug bin is retrieved by a robotic arm and then transported to the input/output (I/O) points by the crane in ADDSs. Fig. 2 provides a visual representation of an ADDS with aisle-captive cranes and bins housing drugs. To fulfil prescription orders, the robotic arm needs to visit the locations with bins that store the requested drugs. In ADDSs, the average visit time to these locations can be minimized by strategically storing highly correlated products in nearby bins.

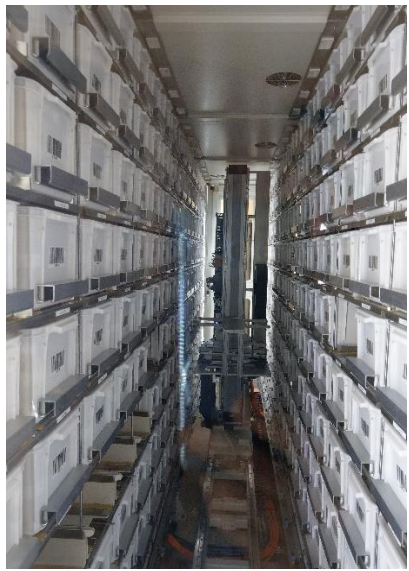

Fig. 2. An example of the internal structure of the ADDS.

Motivated by the above, this study investigates a scattered storage and clustered allocation (SSCA) strategy to optimize the SLAP for drugs in ADDSs. We develop a grouping model based on scattered storage and adopt a sequential alternating (SA) heuristic to cluster the drugs and allocate them in ADDSs. Then, we establish a model for the order-picking problem in a human-robot cooperation environment to verify the effectiveness of the SSCA strategy. The model is validated using a real pharmacy dataset.

The structure of this paper is organized as follows. Section 2 provides a comprehensive review of the literature on the SLAP. In Section 3, we describe and model the SSCA strategy. Moving forward, Section 4 focuses on modelling the order-picking problem in a human-robot cooperation environment. To assess the effectiveness of the proposed strategy, Section 5 presents the numerical experiments and discusses the performance results obtained. Finally, Section 6 concludes the paper by summarizing the main findings and future research.

**2. Literature review**

The storage location assignment problem (SLAP) focuses on optimizing the allocation of products to enhance picking efficiency in order fulfilment problems. This section reviews the relevant literature about the clustering-based storage location assignment problem and order fulfilment problem in a human-robot cooperation environment.

The correlation between products has been studied to develop the clustering-based SLAP. Frazele and Sharp (1989) introduced a correlation ratio to measure the correlation among products and formulated the SLAP as an integer programming problem, which is NP-hard and solved by a two-stage heuristic. Correlation analysis plays a vital role in determining the similarity or correlation between products in the product mix (Bindi et al., 2009). Garfinkel (2005) studied the correlated SLAP where products that are frequently ordered together should be stored together to minimize the multi-zone orders. Jane and Laih (2005) defined similarity as the co-appearance of two items within the order set and utilized it to distribute similar items across different zones, aiming to achieve workload balancing in a synchronized zone order picking system. De Koster et al. (2007) considered the correlation between products and then located the products with a high correlation close to each other. Xiang et al. (2018) formulated a model to decide which product to put in which pod, with the objective of maximizing product similarity. Mirzaei et al. (2021) considered information on the frequency at which products are ordered jointly and introduced an integrated cluster allocation (ICA) approach which clusters products and allocates them to storage locations simultaneously, using both product turnover frequency and correlation based on the assumption that one type of product can be stored on one rack. Zhang et al. (2019) introduced the concept of the demand correlation pattern to describe the correlation among products instead of the pairwise correlation measurement that is commonly employed in the literature, based on which a new model is constructed to address the SLAP. Overall, research on the clustering-based SLAP has emphasized the importance of considering product correlations and demand frequency as influential factors in optimizing the allocation of products within storage systems.

In addition to clustering strategies, researchers have also studied decomposition approaches to solve the SLAP. Frazelle (1989) proposed a two-stage decomposition approach for the storage location assignment problem to minimize order picking time by considering product correlations. Firstly, products are sequentially grouped into clusters, starting with the most frequently ordered items and gradually adding those with the highest correlations until the cluster reaches its capacity. Secondly, clusters with the highest overall popularity are assigned to the closest available locations. Chuang et al. (2012) established a two-stage clustering-assignment model to study the SLAP. In the first stage, products are grouped based on the between-product association. In the second stage, products are assigned to storage locations by the demand frequency of products. They verified that this method is better than the frequency-based assignment method for the order-picking problem. However, the product allocation in the second stage still did not consider the correlation between products, which may not be optimal for assigning the premium locations to products with higher turnover. Yuan et al. (2023) proposed a three-stage approach to optimize the SLAP in ADDS. However, in the third stage, constrained by ADDS's picking mode, they considered the demand frequency of drugs without considering drug correlation. In contrast, the operation of ADDS in this study is different, resulting in distinctions in the methodology of SLAP. In summary, decomposition approaches have been proposed as effective solutions to the SLAP, considering product correlations and popularity. However, there is still room for improvement in incorporating correlation in the SLAP.

  The above studies on SLAP have focused on improving the efficiency of order fulfilment of robots while overlooking the fact that the ADDS is a human-robot cooperation system. Petersen (2005) proposed storage strategies that prioritize placing high-demand items within the 'golden zone' for easier access which refers to the height between the picker's waist and shoulder, which results in significant cost savings in terms of the order fulfilment time. Grosse et al. (2013) investigated the impact of learning and forgetting of a heterogeneous workforce on order picking time and, consequently, on storage assignment decisions. Larco et al. (2016) developed a bi-objective method that makes a trade-off between warehouse efficiency and worker discomfort via storage assignment decisions. Azadeh et al. (2019) emphasized the need for researchers to pay more attention to the interaction between humans and machines in future research. Zhang et al. (2023) examined the relationship between item storage assignment strategies, robots' picking efficiency, and pickers' energy expenditure. Motivated by the above insights, we introduce pharmacists' stochastic picking time in the order-picking process of ADDSs to investigate its impact on prescription order fulfilment.

This paper contributes by introducing a two-stage SSCA strategy for ADDSs which groups drugs based on scattered storage and then allocates them to storage locations by considering both drug demand frequency and correlation. Additionally, to compare the performance of the SSCA strategy, we compare the frequency-based allocation (FA) strategy (formerly referred to as 'method 1') proposed by Yuan et al. (2023), the integrated cluster allocation (ICA) strategy proposed by (Mirzaei et al., 2021), and scattered storage-frequency-based allocation (SFA) strategy. In the numerical experiment, we study the impact of various factors, including order size, cluster size, correlation, and pharmacists' stochastic picking, on the average picking time for ADDSs.

## 3. Problem description and mathematical formulation

This section provides a formal description of the proposed two-stage SSCA strategy for ADDSs. An ADDS consists of a double-sided storage rack, a crane, a robotic arm, a track, and two I/O points, denoted as $O_1$ and $O_2$. The system can be regarded as a unit load automated storage and retrieval system (AS/RS). The storage locations on both sides of the machines can be regarded as grid planes, as shown in Fig. 3. Each storage location is filled with a bin. With the objective of 'storing for picking', the allocation of storage locations for drugs is carried out within the storage area.

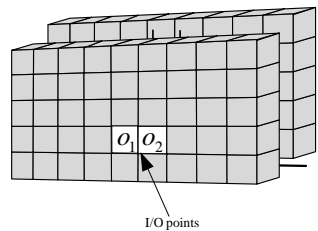

Fig. 3. A view of the ADDS.

For the SSCA strategy, drugs are grouped into ADDSs and each group of drugs are stored in an ADDS. Then in each ADDS, the drugs are allocated to specific locations that consist of drug bins. We divide the SSCA strategy into two stages. Stage I: grouping based on a scattered storage policy. The aim is to group drugs that often appear in the same prescription order based on the correlation of drugs. Stage II: locating based on a clustered allocation policy. For this stage, we assign locations to the drugs within each cluster. Note that all drugs in the prescription orders are available in ADDSs. We make the following assumptions.

- Each storage location (e.g. bin) consists of one location (e.g. one bin). Each bin can store only one type of drug.
- The order history is sufficiently large to accurately capture drug demand frequency and correlation.
- The total storage capacity of ADDSs is sufficient to accommodate all types of drugs.

3.1 Stage I: grouping based on a scattered storage policy

To increase the probability of always having some drugs per ADDS close by, we adopt a scattered storage policy, which means that a type of drug is not stored in a single machine, but scattered over multiple machines. If drugs are scattered over many machines, it becomes advantageous for clustering highly correlated drugs of a prescription order onto a machine within the limited capacity of each machine.

We formulate the 0-1 quadratic mathematical programming model for the first stage of the SSCA strategy, which aims to group drugs by maximizing the total similarity coefficient of the drugs in an ADDS. The notations used are defined as follows.

| | |
|---|---|
| $k$: | The index number of drug, $k \in \{1, 2, \cdots, K\}$, where $K$ is the total number of drugs in the pharmacy. |
| $r$: | The index number of machine, $r \in \{1, 2, \cdots, R\}$. |
| $b_k$: | The storage space (or drug bins) of drug $k$. |
| $f_k$: | The demand frequency of drug $k$. |
| $\eta_k$: | The demand frequency of drug $k$ in one bin, $\eta_k = f_k / b_k$. |
| $Q$: | The total bins of a machine. |

| $x_{k,r}$: | Binary variable. $x_{k,r}=1$, if the drug $k$ is assigned to the machine $r$; and 0, otherwise. |
|---|---|
| $b_{k,r}$: | Integer variable. Number of bins that drug $k$ occupies on the $r$th machine. |
| $S_{k,k'}$: | The Jaccard similarity coefficient value between drug $k$ and $k'$. |

The correlation between drugs $k$ and $k'$ is quantified using the Jaccard similarity coefficient, as depicted by Eq. (1).

$$S_{k,k'} = \gamma_{k,k'} / (\alpha_{k,k'} + \beta_{k,k'} + \gamma_{k,k'}), \tag{1}$$

where $\alpha_{k,k'}$ = the number of prescription orders containing drug $k$ but excluding drug $k'$,

$\beta_{k,k'}$ = the number of prescription orders containing drug $k'$ but excluding drug $k$,

$\gamma_{k,k'}$ = the number of prescription orders containing both drugs $k$ and $k'$.

Note that $S_{k,k'} \in [0,1]$. If $S_{k,k'} = 0$, it means that drugs $k$ and $k'$ do not appear in any prescription order simultaneously. If $S_{k,k'} = 1$, it means that drug $k$ and drug $k'$ always appear in the same prescription order. Then the 0-1 quadratic mathematical programming model is formulated as follows.

$$\max \sum_{k=1}^{K-1} \sum_{k'=k+1}^{K} \sum_{r=1}^{R} x_{k,r} x_{k',r} S_{k,k'} \tag{2}$$

subject to

$$\sum_{r=1}^{R} b_{k,r} = b_k, \forall k \tag{3}$$

$$\sum_{k=1}^{K} b_{k,r} \leq Q, \forall r \tag{4}$$

$$x_{k,r} \leq b_{k,r} \leq Q x_{k,r}, \forall k, r \tag{5}$$

$$x_{k,r} = \begin{cases} 1 & \text{drug } k \text{ is assigned to machine } r, \\ 0 & \text{otherwise.} \end{cases}$$

The term $\sum_{k=1}^{K-1}\sum_{k'=k+1}^{K}\sum_{r=1}^{R} x_{k,r} x_{k',r} S_{k,k'}$ in objective function (2) is to maximize the similarity value of all drugs among all ADDSs. The term $x_{k,r} x_{k',r} S_{k,k'}$ equals $S_{k,k'}$ if and only if drugs $k$ and $k'$ are allocated to the same machine $r$ (i.e., $x_{k,r} = x_{k',r} = 1$), and is 0 otherwise. The objective function (2) shows that as the value of $\sum_{k=1}^{K-1}\sum_{k'=k+1}^{K}\sum_{r=1}^{R} x_{k,r} x_{k',r} S_{k,k'}$ increases, more highly correlated drugs are stored in the same machine. The constraint (3) ensures that $b_k$ bins of drug $k$ are stored in the machines and each drug could be assigned to more than one machine. The constraint (4) restricts that storage space assigned for drugs is no more than that of one machine. The constraint (5) requires that the number of bins that drug $k$ occupies in one machine is no more than the maximum number of bins in this machine and the machine $r$ can serve drug $k$ only if it is selected to store drug $k$.

Given that the clustering problem operates at a strategic level and the schedule is typically stable once determined, the computational time required for solving this model is of secondary significance. To address this, the optimization package CPLEX 12.9 has been used as the solver to compute the model, ensuring a rigorous and accurate approach.

3.2 Stage II: locating based on a clustered allocation policy

Since the locating problem is comparable to the bin packing problem, which is proven to be NP-hard (Garey & Johnson, 1979), it can be shown that this problem is NP-hard too. A sequential heuristic algorithm which was first proposed by Mirzaei et al. (2021) is adopted to optimize the location assignment problem. The sequential heuristic algorithm clusters the drugs by utilizing both the correlation and demand frequency of drugs alternately and then assigns the clusters to the storage locations. And it can be used for solving real-size pharmacy cases.

The sequential heuristic algorithm is detailly described in Algorithm 1. Note that empty bins are preassigned to locations, so we refer to these as locations. Let $L$ be the storage location set and $\Pi$ be the sorted storage location set, based on ascending time to the I/O point. Each location $i$ contains information such as the side of the machine $e_i \in \{1,2\}$, the row index $x_i$, the column index $y_i$, and the drug index $k$. $z_{e_i,x_i,y_i,k}$ is a 0-1 binary variable, and $z_{e_i,x_i,y_i,k} = 1$ indicates that the drug $k$ is assigned to the location $i$ which is on the $e_i$th side, row $x_i$ and column $y_i$ in the ADDS and 0 otherwise. Each storage location is placed in a bin that can store a type of drug. The total number of drug types is $K$. Let $\eta_k$ be the demand frequency of drug $k$. Let $t_i$ be the time from location $i$ to the I/O points. The algorithm assigns the $b_k$ bins of drug $k$ with the highest demand frequency to the storage locations which are closest to the I/O points and let $z_{e_i,x_i,y_i,k} = 1$. Let $s$ be the threshold that determines the correlation between drugs. Drugs with the highest correlation, i.e. $S_{k,k'} \geq s$, are located closest to the drug $k$, and let $z_{e_i',x_i',y_i',k'} = 1$. The capacity of the cluster is $C$, which specifically represents the number of the types of drugs in a cluster. The capacity of the cluster, $C$, and the set of unassigned drugs, $\Lambda$, and the set of unassigned locations, $\Pi$, are updated for each loop. If more than one drug has the highest correlation with $k$, select drug $k'$ with the highest $\eta_k$ closest to drug $k$ (Line

12). If drug $k$ is not correlated with any other drugs, the drug $k'$ with the highest demand frequency is assigned to the location (Line 14). In this way, highly correlated drugs are clustered, while popular drugs are assigned to locations close to the I/O points. This procedure is repeated until all the drugs are assigned to storage locations.

Algorithm 1. (Pseudocode for a sequential alternating (SA) heuristic)

1: input $L$, $C$, $K$, $S_{k,k'}$, $t_i$, $\eta_k$, $i$, $b_k$, $s$, $e_i$, $x_i$, $y_i$, $k$

2: $\Lambda =:$ set of drugs sorted in descending order of $\eta_k$, $k \in K$.

3: $\Pi =:$ set of locations sorted in ascending order of $t_i$, $i \in L$.

4: while $\Lambda \neq \phi$ do

5:    Assign the bins $b_k$ of the first drug $k \in \Lambda$ to the first locations $i$, $z_{e_i,x_i,y_i,k} = 1$.

6:    $\Lambda = \Lambda - \{k\}$.

7:    $\Pi = \Pi - \{i\}$.

8:    $C = C - 1$.

9:    while $C > 0$, $\Lambda \neq \phi$ and $\Pi \neq \phi$ do

10:      select $k' := Arg \max \{S_{k,k'} \mid k' \in \Lambda\}$.

11:      if $k'$ is not unique and $S_{k,k'} \geq s$ then

12:        Assign the drug bins of drug $k'$ with highest $\eta_{k'}$ to locations $i'$. $z_{e'_i,x'_i,y'_i,k'} = 1$.

13:      else if $k'$ is unique and $S_{k,k'} \geq s$ then

14:        Assign the drug bins $b_{k'}$ of drug $k'$ to locations $i'$. $z_{e'_i,x'_i,y'_i,k'} = 1$.

15:      else

16:     Assign the drug bins $b_{k'}$ of the first drug $k'$ in $\Lambda$ to locations $i'$.

$$z_{e_i',x_i',y_i',k'} = 1.$$

17:   end if

18:   $\Lambda = \Lambda - \{k'\}$.

19:   $\Pi = \Pi - \{i'\}$.

20:   $C = C - 1$.

21:   $S_{k,:} = 0$ ($S_{:,k} = 0$), $S_{k',:} = 0$ ($S_{:,k'} = 0$).

22: end while

23: $S_{k,:} = 0$ ($S_{:,k} = 0$).

22: end while

23: return all $z_{e_i,x_i,y_i,k}$.

The optimization result, $z_{e_i,x_i,y_i,k}$, generated by Algorithm 1 is for the location assignment. The second stage of the SSCA strategy is completed.

**4. Order picking problem**

In this section, we focus on the order-picking problem in ADDSs operating under a human-robot cooperation environment. Efficient order fulfilment is crucial in ADDSs. One essential aspect of order fulfilment is the order assignment, which assigns the drugs of the prescription order $q$ to $R$ machines for fulfilment, as shown in Fig. 4.

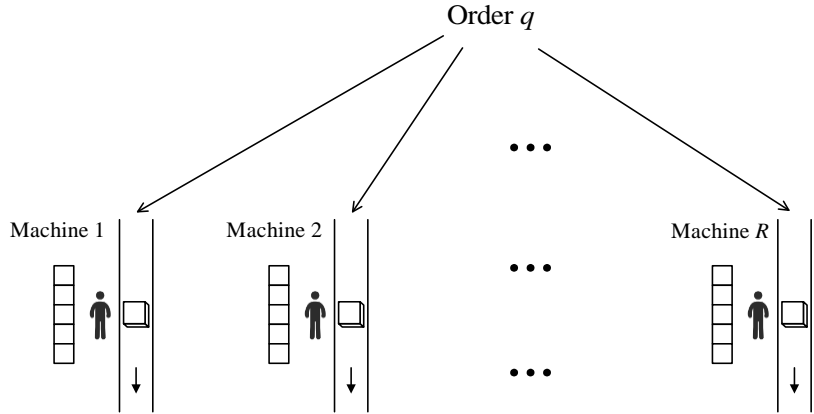

Fig. 4. Order split in ADDSs.

The existing research about order picking is under the assumption that orders have already been assigned to the machines (Boysen et al., 2019). However, we allow splitting orders, i.e. a prescription order may be handled by multiple machines. The model allowing split orders is proved NP-hard by Xie et al. (2021). A split order means that we divide an order into two or more parts for picking (perhaps at different machines). Splitting an order means separating the order into two or more parts (up to the number of drugs in the order). The splitting orders might need additional consolidation time in packing stations and we ignore this time in our study.

Section 4.1 discusses the operation in ADDSs operating under a human-robot cooperation environment, along with the underlying modelling assumptions. In Section 4.2, we establish a double objective integer programming model for the order-picking problem.

4.1 The operation in ADDSs

The operation of ADDSs in a human-robot cooperation environment is as follows. Firstly, prescription information is obtained through the information system. According to the requested information, the robotic arm sends the drug bin from one

I/O point to the storage location firstly, then retrieves the drug bin from the storage area, and sends it to the I/O point finally. The above operation of the robotic arm is a dual command cycle, which is one of the common storage retrieval ways, as shown in Fig. 5, where $i$ and $j$ are two locations in the storage area. Meanwhile, the pharmacist sorts the needed dosage of another type of drug at another I/O point. Suppose the service time $X$ of pharmacists at the I/O point is fixed or normally distributed, i.e. $X \sim N(\mu, \sigma^2)$, where $\mu$ is the mean value and $\sigma$ is the standard deviation. Repeat the above process until all drugs in the prescription order are retrieved.

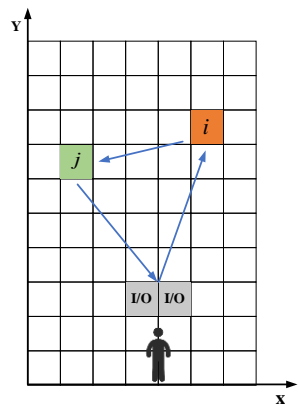

Fig. 5. Dual command cycle operation of the robotic arm.

The time required for the robotic arm to load or unload a drug bin at the I/O point or a storage location is ignored. The time that the robotic arm moves between two I/O points is also ignored. After all drugs of a prescription order are picked, the pharmacist puts them into a container, which is then labelled and delivered to the patient. Prescription orders are released one by one in the sequence they arrive. The robotic arm runs on the middle track and can pick up any bin on both sides. It travels simultaneously in the horizontal and vertical direction at a constant speed $v$ and its acceleration and deceleration are ignored.

Each bin in the ADDS is located on the storage area and its location is denoted by $i$ and $j$, with coordinates $(x_i, y_i)$, where $x_i$ is the row, $y_i$ is the column of the storage location. Let $d_m$ and $d_n$ be the length and height of each bin respectively and all bins are the same. Let $(x_0, y_0)$ denote the coordinates of the I/O points. The travel time from location $i$ to the I/O point is represented by $\max\{|x_i - x_0| \times d_m/v, |y_i - y_0| \times d_n/v\}$. For route $O \to i \to j \to O$ in 错误!未找到引用源。, the dual command travel time, $t_{i,j}$, can be expressed by the following equation (6).

$$t_{i,j} = \max\{|x_i - x_0| \times d_m/v, |y_i - y_0| \times d_n/v\} + \max\{|x_i - x_j| \times d_m/v, |y_i - y_j| \times d_n/v\} \\ + \max\{|x_j - x_0| \times d_m/v, |y_j - y_0| \times d_n/v\}. \quad (6)$$

The picking operations interact between the two I/O points. let $i_k$ denote the location of drug $k$. As shown in Fig. 6, the picking route of the robotic arm is $O_1 \to i_1 \to O_1$, $O_2 \to i_2 \to O_2$, $O_1 \to i_1 \to i_3 \to O_1$, $O_2 \to i_2 \to i_4 \to O_2$, $\cdots$, $O_1 \to i_{K-1} \to O_1$, $O_2 \to i_K \to O_2$, where $K$ is the total types of drugs in a prescription order. There are $K-2$ dual command operations if $K \geq 2$.

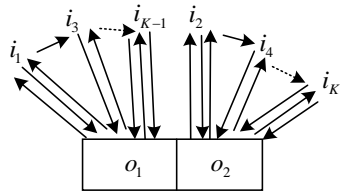

Fig. 6. Picking route.

4.2 The integer programming model for the order picking problem

We establish a double objective integer programming model for the order-picking problem to minimize the number of machines visited in the prescription orders while keeping the least average picking time for a prescription order. The notations used are as follows.

| | |
|---|---|
| $k$: | The drug index in a prescription order, $k \in \{1, 2, \ldots, K_q\}$, where $K_q$ is the number of drug types in the order $q$. |
| $X$: | Service time of pharmacists at the I/O point. |
| $r$: | Machine index, $r \in \{1, 2, \ldots, R\}$. |
| $M_k$: | The location set of bins for drug $k$. $\Phi_r = M_1 \cup M_2 \cup \cdots \cup M_{K_q}$. |
| $s_j$: | Remaining stock at the location $j$. |
| $a_k$: | The dosage of drug $k$ in the prescription order. |
| $t_{i,j}$: | The dual command travel time of the robotic arm from location $i$ to $j$, $i \in \Phi_r$, $j \in \Phi_r$. |
| $p$: | The index of I/O points. $p \in \{1, 2\}$. That is, there are two reserved spaces as the I/O points in an ADDS. |
| $x^r_{i,j,p}$: | 1, if the location $j$ is visited immediately after $i$ in the route from the $p$th I/O point in the machine $r$; 0, otherwise. |
| $z_{q,r}$: | 1, if order $q$ is assigned to machine $r$; 0, otherwise. |
| $z_{k,q,r}$: | 1, if drug $k \in K$ of order $q$ is assigned to machine $r \in R$; 0, otherwise. |

Let

$$\tilde{t}_{1,1} = t_{i,i} I\{\exists\, j, x^r_{i,j,1} = 1, \text{ and } x^r_{j',i,1} = 0 \text{ for } \forall j', r\}, \tag{7}$$

$$\tilde{t}_{2,2} = t_{i,i} I\{\exists\, j, x^r_{i,j,2} = 1, \text{ and } x^r_{j',i,2} = 0 \text{ for } \forall j', r\}, \tag{8}$$

$$\tilde{t}_{K_q-1, K_q-1} = t_{j,j} I\{\exists\, i, x^r_{i,j,p} = 1, \text{ and } x^r_{j,i',p} = 0 \text{ for } \forall i', r, p = (K_q \bmod 2) + 1\}, \tag{9}$$

$$\tilde{t}_{K_q,K_q} = t_{j,j} I\left\{\exists i, x^r_{i,j,p} = 1, \text{ and } x^r_{j,i',p} = 0 \text{ for } \forall i', r, p = K_q \bmod 2 + 2\right\}, \quad (10)$$

where $I\{x\}$ is an indicator function and $I\{x\} = \begin{cases} 1, & \text{if } x \text{ is satisfied,} \\ 0, & \text{others.} \end{cases}$

For example,

$$I\left\{\exists j, x^r_{i,j,1} = 1, \text{ and } x^r_{j',i,1} = 0 \text{ for } \forall j', r\right\} = \begin{cases} 1, & \text{if } \exists j, x^r_{i,j,1} = 1, \text{ and } x^r_{j',i,1} = 0 \text{ for } \forall j', r, \\ 0, & \text{others.} \end{cases}$$

The model of double objective 0-1 integer programming model is proposed as follows.

$$\min \sum_q \sum_r z_{q,r} \quad (11)$$

$$\min \sum_{r=1}^{R} (\tilde{t}_{1,1} + \max(X, \tilde{t}_{2,2})) + \sum_{p=1}^{2} \sum_{\substack{j=1 \\ j \neq i}}^{|\Phi_r|} \sum_{i=1}^{|\Phi_r|} E\left[\max(X, t_{i,j})\right] x^r_{i,j,p} + \max(X, \tilde{t}_{K_q-1, K_q-1}) + \tilde{t}_{K_q, K_q}) \quad (12)$$

subject to

$$z_{q,r} \geq z_{k,q,r}, \forall q, k, r \quad (13)$$

$$\sum_k z_{k,q,r} \geq z_{q,r}, \forall q, r \quad (14)$$

$$\sum_r z_{k,q,r} = 1, \forall q, k \quad (15)$$

$$\sum_{\substack{j \in \Phi_r \\ j \neq i}} \sum_{i \in \Phi_r} x^r_{i,j,p} = \sum_k z_{k,q,r}, \forall q, r \quad (16)$$

$$a_k x^r_{i,j,p} \leq s_j, \forall i, j, k, r, p \quad (17)$$

$$\sum_p \sum_{i \in \Phi_r} x^r_{i,j,p} \leq 1, \forall j, r \quad (18)$$

$$\sum_p \sum_{j \in \Phi_r} x^r_{i,j,p} \leq 1, \forall i, r \quad (19)$$

$$\sum_p \sum_{i \in \Phi_r} x^r_{i,j,p} = \sum_p \sum_{i \in \Phi_r} x^r_{j,i,p}, \forall j, r \tag{20}$$

$$\sum_{\substack{j \in M_k \\ j \neq i}} \sum_{i \in M_k} x^r_{i,j,p} = 0, \forall p, k, r \tag{21}$$

$$\sum_{j \in M_k} \sum_{i \notin M_k} x^r_{i,j,p} = 1, \forall r, k, p \tag{22}$$

$$\left| \sum_{\substack{j \in \Phi_r \\ j \neq i}} \sum_{i \in \Phi_r} x^r_{i,j,1} - \sum_{\substack{j \in \Phi_r \\ j \neq i}} \sum_{i \in \Phi_r} x^r_{i,j,2} \right| \leq 1, \forall r \tag{23}$$

$$u_i - u_j + |\Phi_r| x^r_{i,j,p} \leq |\Phi_r| - 1, 1 < i \neq j \leq |\Phi_r| \tag{24}$$

$$x^r_{i,j,p} = 0,1, \forall i, j \in \Phi_r, i \neq j, p \in \{1,2\}, r \in \{1,2,\cdots,R\} \tag{25}$$

In the above model, the objective function (12) is to minimize the average picking time for the order picking in a human-machine interaction environment, where

$$E\left[\max(X, t_{i,j})\right] = t_{i,j} \int_0^{t_{i,j}} f(x) dx + \int_{t_{i,j}}^{+\infty} x f(x) dx. \tag{26}$$

Constraint (13) shows that the drugs of an order don't have to be assigned to the same machine anymore. Constraint (14) ensures that an order can only be assigned to a machine if at least one drug of the order is assigned to that machine. Constraint (15) ensures that each drug of the order can be assigned to one machine. Constraint (16) ensures that there are $\sum_k z_{k,q,r}$ requests performed by dual command cycle operations, which ensures that all drugs of the order $q$ in the $r$th machine can be handled by the machine. Constraint (17) shows that the remaining stock at location $j$ meets the dosage of the drug $k$ when location $j$ is visited, i.e. $x^r_{i,j,p} = 1$. Constraints (18) and

(19) limit that each location can be visited at most once. Constraint (20) indicates that the number of input arcs at each point is equal to the number of output arcs for the route of the robotic arm. Constraint (21) avoids the drug being repeatedly picked in the prescription order. Constraint (22) ensures that the robotic arm operations from one drug set to another drug set. Affected by the working feature of the robotic arm in ADDSs, constraint (23) ensures that the robotic arm works alternately between two I/O points. The sub-tour elimination constraints are described by constraint (24), where $u_i \ (i=1,2,\cdots,|\Phi_r|)$ are arbitrary real numbers (Miller et al., 1960). The model above is programmed in Matlab and solved using Cplex 12.9.

**5. Numerical analysis**

To evaluate the performance of the two-stage SSCA strategy, we use a dataset with patients' demands for a pharmacy, available from Changi General Hospital in Singapore. This dataset consists of about 10000 prescription orders and 488 types of drugs. There are $9 \times 16$ positions to locate bins on each side of ADDSs. The positions $(8,6)$ and $(9,6)$ on side 1 are two I/O points. The horizontal speed and the vertical speed of the robotic arm are $v = 0.1486 m/s$. And the length of each storage bin is $d_n = 0.168m$ and the height is $d_m = 0.275m$.

In Section 5.1, we illustrate the details of comparative SLAP strategies. Section 5.2 discusses the performance measures under different SLAP strategies. Section 5.3 and Section 5.4 show the average picking time of order picking problem by four strategies in a fully automated environment and a human-robot cooperation environment respectively.

5.1 Comparative SLAP strategies

This section provides a comprehensive description of the two-stage SSCA strategy in ADDSs, the frequency-based allocation (FA) strategy, the integrated cluster allocation (ICA) strategy, and the scattered storage and frequency-based allocation (SSFA) strategy. Each of these strategies is detailed below, highlighting their distinct characteristics. Table 1 presents a clear overview of the main differences between these strategies.

SSCA (our strategy): The first stage of this strategy is to group drugs based on a scattered storage policy, which is described in detail in Section 3.1. The second stage is to locate drugs based on a clustered allocation policy by utilizing both the correlation and demand frequency of drugs alternately, which is described in detail in Section 3.2.

FA: The first stage of this strategy involves a general grouping model without scattered storage. In the second stage, frequency-based allocation is employed for drug locating. This approach was originally proposed by Yuan et al. (2023).

ICA: The first stage is the same as that of the FA strategy, followed by the second stage, which is the same as the one of the SSCA strategy. It was proposed by Mirzaei et al. (2021).

SSFA: The first stage is the same as that of the SSCA strategy, and the second stage is the same as that of the FA strategy. We refer to this approach as a scattered storage and frequency-based allocation (SSFA) strategy. This strategy primarily serves to validate the effectiveness of the second stage of our proposed strategy.

Table 1.

The main differences between FA, ICA, SSFA and SSCA strategies.

|  | FA | ICA | SSFA | SSCA |
|---|---|---|---|---|
| Scattered storage | - | - | √ | √ |
| Clustered allocation | - | √ | - | √ |

5.2 The performance measures under four strategies

In this section, we primarily focus on discussing key parameters that significantly impact the optimization process. These parameters include the size of the prescription order, the size of clusters, and the threshold of the correlation between drugs. The values of the above parameters are presented in Table 2, where $\sigma = 0$, 2, and 5 represent three variation scenarios of pharmacists' sorting time: fixed, medium variability, and high variability respectively.

Table 2

Parameters.

| Parameters | Symbols | Description | Range |
|---|---|---|---|
| Order size | $|K|$ | Number of drugs | [1, 5], [6, +∞] |
| Clusters | $C$ | Number of drug types in a cluster | 2-14 |
| Threshold | $s$ | The threshold of the correlation between drugs | 0.001, 0.01, 0.1 |
| Time | $X$ | Sorting time of pharmacists | $\mu = 0, 5, 10, 15,$ $\sigma = 0, 2, 5$ |

To examine the impact of order size on the system efficiency of ADDSs, a statistical analysis is conducted to the prescription orders of 10 days. Fig. 7 shows that a majority of prescription orders are in the range of [1,5] in terms of prescription order size, accounting for nearly 90% of the total.

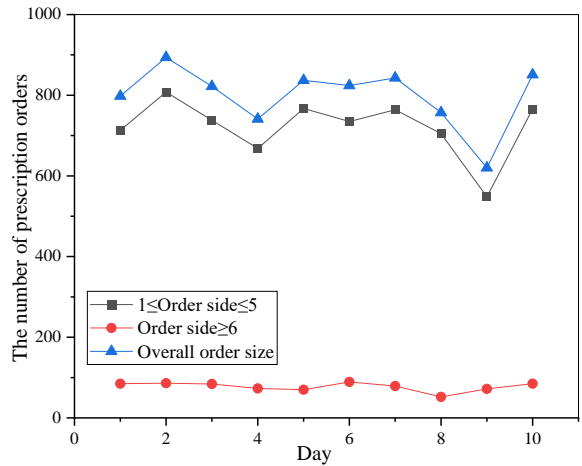

Fig. 7. The number of different prescription order size.

If one order is split and picked by two or more machines, we call it a cross-machine prescription order. The probability of cross-machine prescription orders is examined and the results are shown in Fig. 8. Fig. 8 shows that FA and ICA strategies cause a higher probability of cross-machine prescription orders compared to SSFA and SSCA strategies. The improvement of SSFA and SSCA strategies in the probability of cross-machine prescription orders is 1.56%-4.35%. This reaches our expectations, primarily due to the benefits of the scattered storage in increasing the likelihood of grouping correlated drugs together in the limited-capacity machine.

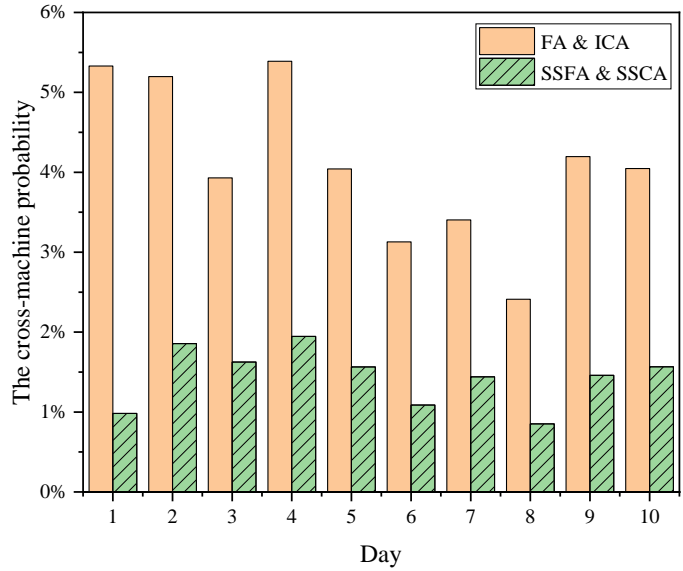

Fig. 8. The cross-machine probability under different policies.

Let the cross-machine time of one prescription order be 60 seconds. We investigate the impact of the cluster size and threshold of the correlation between drugs between drugs on the average picking time of prescription orders. The results are shown in Fig. 9 and Fig. 10.

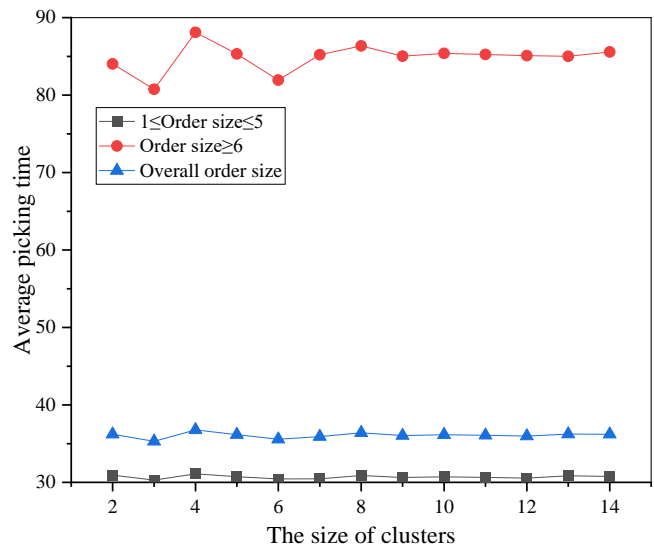

Fig. 9. The impact of the cluster size on the average picking time.

Fig. 9 illustrates the influence of cluster size on the average picking time for the SSCA strategy. The results reveal that the cluster size has a significant impact on the large order size (order size≥6) and the optimal performance is achieved when the cluster size is 3, where the average picking time is 80.74s. For the case of larger order sizes (order size ≥ 6), the maximum time is 88.10s when the cluster size is 4. Interestingly, a turning point in the average picking time is observed for cluster sizes that are multiples of 3, such as 3, 6, and 9. The variation of the average picking time of drugs is not directly proportional to the size of the cluster. This observation may be related to the fact that the average number of drugs per prescription is 3.

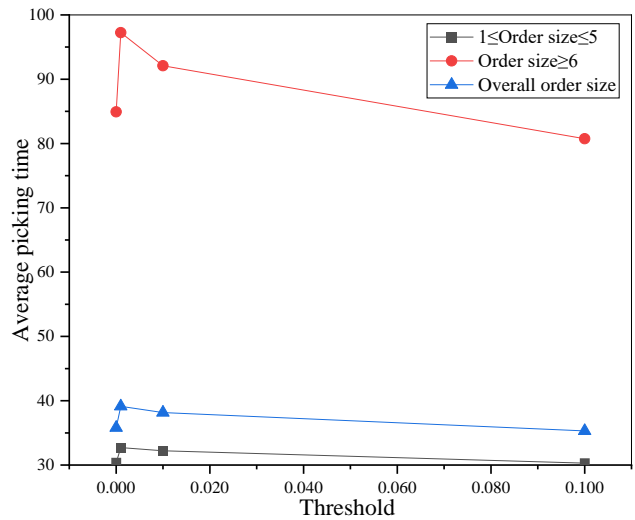

Fig. 10. The impact of the threshold of the correlation between drugs.

Fig. 10 investigates the impact of the threshold of the correlation between drugs on the average picking time of prescription orders for the SSCA strategy, specifically focusing on the scenario where cluster size is 3. The results demonstrate that as the threshold increases, the average picking time is reduced for any case of order size. This is due to the enhanced threshold of the correlation between drugs, resulting in a higher possibility of drug grouping. Therefore, it reduces the possibility of cross-machine picking for prescription orders.

We further examine variations in order size. Notably, the most significant improvement in average picking time is observed for larger order sizes (order size $\geqslant$ 6) and the range of changes in average picking time for three order sizes is 0.81s, 7.36s, and 1.46s, respectively. Since prescription orders generally contain a small number of drugs, with most falling within the range of [1, 5], the improvement in average picking time is mainly influenced by the smaller prescription order size. Consequently, the improvement of the overall order size prescription orders is not as significant as the large order size (order size $\geqslant$ 6).

Combining the observations from Fig. 9 and Fig. 10, it can be inferred that the average picking time is influenced by both cluster size and the threshold of the correlation between drugs within the clusters.

## *5.3 Comparison of four strategies in a fully automated environment*

We compare the SSCA strategy with the FA, ICA, and SSFA strategies in a fully automated environment for ADDSs. That is, there are no human picking factors, i.e., the picking time of pharmacists is zero. The results are shown in Fig. 11.

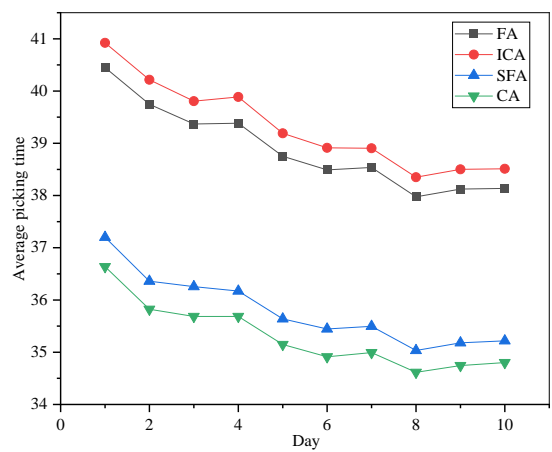

Fig. 11. Average picking time of the FA, ICA, SSFA and SSCA strategies.

Fig. 11 illustrates the average picking time for prescription orders per day. We find that the average picking time satisfies SSCA < SSFA < ICA < FA. Notably, both

SSFA and SSCA strategies perform better than ICA and FA strategies, primarily due to the effectiveness of the scattered storage in the first stage of the SSCA and SSFA strategies. The grouping based on a scattered storage policy directly determines the effectiveness, whether it is the frequency-based allocation or the clustered allocation for location assignment. When more drugs in an order are allocated to a machine, the scattered storage policy proves to be more effective. Conversely, the frequency-based allocation is better when the number of drugs in an order assigned to each machine is lower.

*5.4 Comparison of four strategies in a human-robot cooperation environment*

We further investigate the impact of the stochastic picking time of pharmacists on the system performance. Table 3 shows the results for the $(\mu,\sigma)$ defined in Table 2. The first column in Table 3 is the parameter $(\mu,\sigma)$. The average picking time of prescription orders are compared among the FA, ICA, SSFA, and SSCA strategies.

Table 3.

The average picking time of four strategies under different $(\mu,\sigma)$.

| $(\mu,\sigma)$ | FA | ICA | SFA | CA |
| --- | --- | --- | --- | --- |
| (0,0) | 38.90 | 39.32 | 35.80 | 35.30 |
| (5,0) | 42.05 | 42.46 | 38.77 | 38.35 |
| (5,2) | 42.30 | 42.69 | 39.05 | 38.66 |
| (5,5) | 43.59 | 43.92 | 40.43 | 40.06 |
| (10,0) | 48.01 | 48.18 | 44.88 | 44.65 |
| (10,2) | 48.27 | 48.46 | 45.16 | 44.92 |

| (10,5) | 49.57 | 49.78 | 46.52 | 46.26 |
| (15,0) | 57.20 | 57.27 | 54.41 | 54.31 |
| (15,2) | 57.40 | 57.45 | 54.56 | 54.45 |
| (15,5) | 58.18 | 58.26 | 55.31 | 55.16 |

From Table 3, we can find that the SSCA strategy performs better than the FA, ICA, and SSFA strategies for all cases. For any fixed $\mu$, the average picking time for the four strategies increases with $\sigma$. Similarly, for any fixed $\sigma$, the average picking time for the four strategies also increases with $\mu$. This also indicates that 'humans' influence the picking efficiency greatly in a human-robot cooperation environment.

## 6. Conclusion and future research

In automated drug dispensing systems (ADDSs), the assignment of drug storage locations significantly affects the order-picking process. Specifically, as ADDSs operate as human-robot collaborative systems, exploring the impact of stochastic picking times by pharmacists on the efficiency of order picking is crucial. Therefore, this study proposes a two-stage scattered storage and clustered allocation (SSCA) strategy to optimize the storage location assignment problem (SLAP). Furthermore, within the human-robot cooperation environment, an order-picking model is established, expanding on previous models by allowing order splitting. In our experimental pharmacy order instances, we investigate the influence of order size, cluster size, and the correlation threshold between drugs on cross-machine probability and average picking time of prescription orders. To assess the performance of the SSCA strategy, we compare it with commonly used SLAP strategies in the most recent literature, namely FA, ICA, and SSFA. The results demonstrate that the SSCA strategy has the minimum cross-machine probability and the minimum average picking time for each prescription order. In particular, within a human-robot cooperation environment, we analyse the impact of stochastic picking time associated with pharmacists on average picking time under the SLAP strategies. The SSCA strategy consistently exhibits superior performance across various parameter settings, thus enhancing order-picking efficiency in ADDSs.

Future research may study a dynamic storage assignment strategy to better utilize the storage space for ADDSs. Metaheuristics methods may be interested in finding better product allocations that reduce the optimality gap and solve larger instances. Additional analysis may study the effect of batching orders, which can increase the correlation between the orders and, consequently, the effectiveness of the SSCA strategy.

## References

Azadeh, K., De Koster, R.& Roy, D. (2019). Robotized and automated warehouse systems: Review and recent developments. *Transportation Science*, *53*(4), 917-945. https://doi.org/10.1287/trsc.2018.0873.

Bindi, F., Manzini, R., Pareschi, A.& Regattieri, A. (2009). Similarity-based storage allocation rules in an order picking system: An application to the food service industry. *International Journal of Logistics Research and Applications*, *12*(4), 233-247. https://doi.org/10.1080/13675560903075943.

Boysen, N., Briskorn, D.& Emde, S. (2017). Parts-to-picker based order processing in a rack-moving mobile robots environment. *European Journal of Operational Research*, *262*(2), 550-562. https://doi.org/10.1016/j.ejor.2017.03.053.

Boysen, N., de Koster, R.& Weidinger, F. (2019). Warehousing in the e-commerce era: A survey. *European Journal of Operational Research*, *277*(2), 396-411. https://doi.org/10.1016/j.ejor.2018.08.023.

Chuang, Y.-F., Lee, H.-T.& Lai, Y.-C. (2012). Item-associated cluster assignment model on storage allocation problems. *Computers & Industrial Engineering*, *63*(4), 1171-1177. https://doi.org/10.1016/j.cie.2012.06.021.

De Koster, R., Le-Duc, T.& Roodbergen, K. J. (2007). Design and control of warehouse order picking: A literature review. *European Journal of*

*Operational Research*, *182*(2), 481-501. https://doi.org/10.1016/j.ejor.2006.07.009.

Frazele, E. A.& Sharp, G. P. (1989). Correlated assignment strategy can improve any order-picking operation. *Industrial Engineering*, *21*(4), 33–37.

Frazelle, E. H. (1989). Stock location assignment and order picking productivity. *Georgia Institute of Technology*, PhD dissertation.

Garey, M. R.& Johnson, D. S. (1979). *Computers and intractability: A guide to the theory of np-completeness*. W.H. Freeman.

Garfinkel, M. (2005). Minimizing multi-zone orders in the correlated storage assignment problem. *Georgia Institute of Technology*, PhD dissertation.

Grosse, E. H., Glock, C. H.& Jaber, M. Y. (2013). The effect of worker learning and forgetting on storage reassignment decisions in order picking systems. *Computers & Industrial Engineering*, *66*(4), 653-662. https://doi.org/10.1016/j.cie.2013.09.013.

Gu, J., Goetschalckx, M.& McGinnis, L. F. (2007). Research on warehouse operation: A comprehensive review. *European Journal of Operational Research*, *177*(1), 1-21. https://doi.org/10.1016/j.ejor.2006.02.025.

Hausman, W. H., Schwarz, L. B.& Graves, S. C. (1976). Optimal storage assignment in automatic warehousing systems. *Management Science*, *22*(6), 629-638. https://doi.org/10.1287/mnsc.22.6.629.

Jane, C.-C.& Laih, Y.-W. (2005). A clustering algorithm for item assignment in a synchronized zone order picking system. *European Journal of Operational Research*, *166*(2), 489-496. https://doi.org/10.1016/j.ejor.2004.01.042.

John J. Bartholdi, I.& Hackman, S. T. (2019). *Warehouse & distribution science release 0.98.1*. https://www.warehouse-science.com/.

Larco, J. A., de Koster, R., Roodbergen, K. J.& Dul, J. (2016). Managing warehouse efficiency and worker discomfort through enhanced storage assignment decisions. *International Journal of Production Research*, *55*(21), 6407-6422. https://doi.org/10.1080/00207543.2016.1165880.

Miller, C. E., Tucker, A. W.& Zemlin, R. A. (1960). Integer programming formulation of traveling salesman problems. *Journal of the ACM*, *7*(4), 326–329. https://doi.org/10.1145/321043.321046.

Mirzaei, M., Zaerpour, N.& de Koster, R. (2021). The impact of integrated cluster-based storage allocation on parts-to-picker warehouse performance. *Transportation Research Part E: Logistics and Transportation Review*, *146*(C). https://doi.org/10.1016/j.tre.2020.102207.

Petersen, C. G. (2005). Improving order picking performance utilizing slotting and golden zone storage. *International Journal of Operations & Production Management*, *25*, 997–1012. https://doi.org/10.1108/01443570510619491.

Roodbergen, K. J.& Vis, I. F. A. (2009). A survey of literature on automated storage and retrieval systems. *European Journal of Operational Research*, *194*(2), 343-362. https://doi.org/10.1016/j.ejor.2008.01.038.

Weidinger, F.& Boysen, N. (2018). Scattered storage: How to distribute stock keeping units all around a mixed-shelves warehouse. *Transportation Science*, *52*(6),

1412-1427. https://doi.org/10.1287/trsc.2017.0779.

Xiang, X., Liu, C.& Miao, L. (2018). Storage assignment and order batching problem in kiva mobile fulfilment system. *Engineering Optimization*, *50*(11), 1941-1962. https://doi.org/10.1080/0305215x.2017.1419346.

Xie, L., Thieme, N., Krenzler, R.& Li, H. (2021). Introducing split orders and optimizing operational policies in robotic mobile fulfillment systems. *European Journal of Operational Research*, *288*(1), 80-97. https://doi.org/10.1016/j.ejor.2020.05.032.

Yang, P., Zhao, Z.& Guo, H. (2020). Order batch picking optimization under different storage scenarios for e-commerce warehouses. *Transportation Research Part E: Logistics and Transportation Review*, *136*, Article 101897. https://doi.org/10.1016/j.tre.2020.101897.

Yuan, M., Zhao, N., Wu, K.& Chen, Z. (2023). The storage location assignment problem of automated drug dispensing machines. *Computers & Industrial Engineering*, *184*, Article 109578. https://doi.org/10.1016/j.cie.2023.109578.

Zhang, J., Zhang, N., Tian, L., Zhou, Z.& Wang, P. (2023). Robots' picking efficiency and pickers' energy expenditure: The item storage assignment policy in robotic mobile fulfillment system. *Computers & Industrial Engineering*, *176*, Article 108918. https://doi.org/10.1016/j.cie.2022.108918.

Zhang, R.-Q., Wang, M.& Pan, X. (2019). New model of the storage location assignment problem considering demand correlation pattern. *Computers & Industrial Engineering*, *129*(Mar.), 210-219. https://doi.org/10.1016/j.cie.2019.01.027.

Zhang, Y. (2016). Correlated storage assignment strategy to reduce travel distance in order picking. *IFAC-PapersOnLine*, *49*(2), 30-35. https://doi.org/10.1016/j.ifacol.2016.03.006.